\documentclass[draft]{agujournal2019}
\usepackage{url} 
\usepackage[T1]{fontenc}
\usepackage{lineno}
\usepackage[inline]{trackchanges} 
\usepackage{soul}
\usepackage{amsmath,bm,amssymb}

\draftfalse

\journalname{Geophysical Research Letters}

\begin{document}

\title{Observational evidence for solar wind proton heating by ion-scale turbulence}

\authors{G. Q. Zhao\affil{1,2}, Y. Lin\affil{3}, X. Y. Wang\affil{3}, D. J. Wu\affil{4}, H. Q. Feng\affil{1}, Q. Liu\affil{1}, A. Zhao\affil{1}, and H. B. Li\affil{1}}

 \affiliation{1}{Institute of Space Physics, Luoyang Normal University, Luoyang, China}
 \affiliation{2}{Henan Key Laboratory of Electromagnetic Transformation and Detection, Luoyang, China}
 \affiliation{3}{Physics Department, Auburn University, Auburn, USA}
 \affiliation{4}{Purple Mountain Observatory, CAS, Nanjing, China}

\correspondingauthor{G. Q. Zhao}{zgqisp@163.com}

\begin{keypoints}
\item Nearly collisionless solar wind turbulence at ion scales is investigated with 11 years of in-situ data
\item Correlations between the spectral index and magnetic helicity, and between the proton temperature and turbulent energy are revealed
\item A scenario for the solar wind turbulence and heating at ion scales is proposed
\end{keypoints}


\begin{abstract}
Based on in-situ measurements by Wind spacecraft from 2005 to 2015, this letter reports for the first time a clearly scale-dependent connection between proton temperatures and the turbulence in the solar wind. A statistical analysis of proton-scale turbulence shows that increasing helicity magnitudes correspond to steeper magnetic energy spectra. In particular, there exists a positive power-law correlation (with a slope $\sim 0.4$) between the proton perpendicular temperature and the turbulent magnetic energy at scales $0.3 \lesssim k\rho_p \lesssim 1$, with $k$ being the wavenumber and $\rho_p$ being the proton gyroradius. These findings present evidence of solar wind heating by the proton-scale turbulence. They also provide insight and observational constraint on the physics of turbulent dissipation in the solar wind.
\end{abstract}

\section*{Plain Language Summary}
The solar wind is a tenuous magnetized plasma that serves as a natural laboratory of nearly collisionless turbulence. It is streaming outward from the Sun and has temperatures much higher than those from a spherically expanding ideal gas, implying a heating process must occur. The heating has been extensively discussed in past decades, but still not well understood. Based on in-situ measurements, we reveal a scale-dependent connection between proton temperatures and the turbulence with enhanced magnetic helicity signature. The connection is particularly strong for the proton temperature in the direction perpendicular to the background magnetic field. These observations provide implications for the physics of turbulent dissipation and heating in a collisionless plasma.

\section{Introduction}
It is well known that the solar wind is pervaded with magnetic fluctuations that are inherently turbulent. The fluctuation energy appears as a power-law spectrum over a broad range of spacial scales \cite{ale13p01,bru13p02}. The energy-containing range indicates very large scales with frequencies $f < 10^{-4}$ Hz in the spacecraft reference frame. The energy spectrum in this range goes as $\sim f^{-1}$, and is interpreted as uncorrelated large-scale Alfv\'en waves that provide a source of energy. The inertial range represents intermediate scales with spacecraft frame frequencies $10^{-4} \lesssim f \lesssim 10^{-1}$ Hz. The spectrum follows a $f^{-5/3}$ law that is dominated by the Kolmogorov cascade. At small scales comparable to the proton gyroradius or inertial length, the spectrum becomes complicated with various spectral indices between $-2$ and $-4$ \cite<e.g.,>[]{sah13p15,bru14p15}, frequently lower than the classic index $-7/3$ predicted by strong-turbulence theory for dispersive cascade \cite{gal06p21,wud20}. The spectral steepening at the small scales are often interpreted as the occurrence of turbulent dissipation due to cyclotron damping \cite{gar04p05,smi12p08}, Landau damping \cite{hej15p76,how18p05}, stochastic heating \cite{joh01p21,cha10p03}, or plasma coherent structures such as current sheets and magnetic vortices \cite{per12p01,wan19p22}.

Various studies were conducted to explore the nature of the proton-scale turbulence in the solar wind, which is fundamentally important to the understanding of the dissipation and heating. Two models were proposed theoretically, including kinetic Alfv\'en wave (KAW) and whistler wave turbulence models \cite{gar09p05}. KAW turbulence model was suggested by a large body of studies on the basis of observations \cite<e.g.,>[]{bal05p02,sah09p02,how10p49,hej12p08,pod13p29} and simulations \cite{how08p04,gro18p01}. Among them, several studies used non-zero magnetic helicity to diagnose the presence of KAWs in the turbulence \cite{how10p49,hej12p08,pod13p29}. The KAW turbulence would heat the solar wind through cyclotron/Landau damping or stochastic heating, while it is under debate which process is dominant \cite{par15p13,ise19p63}.

The issue on the heating of the solar wind is complicated because of the complex roles of double-adiabatic expansion, global heat flux, ion differential flows, Coulomb collisions, local wave action, microinstabilities, and turbulence \cite<see a recent review by>[]{ver19p05}. Various correlations in the solar wind have been reported. Proton temperatures are found to be correlated with the solar wind speed, and faster solar wind usually corresponds to higher proton temperatures \cite{mar82p52,new98p53}. Steeper spectra at proton scales are also linked to faster solar wind \cite{bru14p15}. The faster solar wind tends to be more imbalanced with greater wave energy flux anti-sunward than sunward \cite{tuc95p01}. These correlations with the solar wind speed are possibly a consequence of different source populations of the solar wind \cite{sch06p51}. There is also a correlation between the temperature ratio of alpha particles to protons and the alpha$-$proton differential flow \cite{kas08p03}. This has been interpreted as evidence of solar wind heating by field-aligned ion-cyclotron waves \cite{gar06p05,kas13p02}. A link between inertial-range intermittency and proton temperature enhancement has also been reported, as evidence of nonuniform heating driven by magnetohydrodynamic turbulence \cite{osm11p11,osm12p02}.

In this letter, we report a new correlation between the spectral steepness and magnetic helicity for nearly collisionless solar wind turbulence at proton scales. The correlation exists even within a very narrow range of the solar wind speed. Moreover, it is shown that the proton perpendicular temperature is correlated with the turbulent magnetic energy at proton scales. A higher magnetic helicity favors a better correlation of the temperature with the magnetic energy. These findings will provide improved understanding of the ion-scale turbulence and heating in the solar wind.

\section{Data and Analysis Methods}
The data used in this letter are from the Wind spacecraft. The magnetic
field data are from the Magnetic Field Investigation instrument and have a cadence of 0.092 s,  \cite{lep95p07}. The plasma data are from the Solar Wind Experiment instrument working at
a cadence of 92 s \cite{ogi95p55}. The proton temperatures are produced via a nonlinear-least-squares bi-Maxwellian fit of ion spectrum from the Faraday cup \cite{kas06p05}. The data set is surveyed from 2005 to 2015 through dividing the long time series into consecutive and overlapping time segments. In each segment with data available, the magnetic energy spectrum is obtained via standard fast Fourier transform technique. The plasma parameters, including the proton density $N_p$, perpendicular and parallel thermal velocities $w_{{\perp}p}$ and $w_{{\parallel}p}$, and bulk velocity ${\bm V}_p$, are given as average values over the time segment, where $\perp$ and $\parallel$ are with respect to the ambient field ${\bm B}_0$. Each time segment has a span of 200 s, a compromise time interval to reduce the averaging effects for plasma parameters while to cover the magnetic spectrum of interest with an appropriate resolution simultaneously. An overlap is set to be 100 s, comparable to the plasma data cadence. The Coulomb collisional age $A_c$ is calculated, which is the ratio of the transit time of the solar wind to the collision timescale \cite{liv86p45}. Segments with $A_c < 0.1$ are selected for samples with negligible collision effects. This selection means that the faster solar wind would be chosen in principle. The selected segments have a median of the solar wind speed being 519 km s$^{-1}$, while it is 388 km s$^{-1}$ for all segments.
It is also required that the angle between ${\bm B}_0$ and ${\bm V}_p$ is in the range from $60^\circ$ to $120^\circ$ to reduce possible heating/cooling effects due to the alpha$-$proton differential flow, which could be strong when ${\bm B}_0$ and ${\bm V}_p$ are quasi-parallel \cite{zha20p14,zha19p60}. In total about $2.6 \times 10^5$ time segments satisfying these constraints are selected, and most ($95\%$) of them have proton parallel beta in the range from 0.1 to 3.

The normalized reduced magnetic helicity is considered as an indicator of the presence of polarized waves in this letter. The helicity is a measure of the spatial handedness of the magnetic field, and has been widely used to identify wave modes in the solar wind for nearly parallel propagating cyclotron waves \cite{zha18p15,zha19p75,woo19p53} or oblique KAWs  \cite{how10p49,hej11p85}. The helicity refers to the fluctuating helicity in spectral form and is defined as ${H}_m({\bm k}) \equiv {\bm A}({\bm k}){\cdot}{\bm B}^{\ast}({\bm k})$, where ${\bm A}$ is the fluctuating magnetic vector potential, ${\bm B}$ is the fluctuating magnetic field, the asterisk represents the complex conjugate of the Fourier coefficients, and ${\bm k}$ is the wave vector \cite{mat82p11,woo19p53}. For the magnetic field measured by a single spacecraft, only the reduced helicity spectrum is available, which can be written as ${H}_m^r({k_1}) = {2\mathrm{Im}[{B_2}({k_1}){\cdot}{B_3}^{\ast}({k_1})]}/{k_1}$ \cite{mat82p11,woo19p53},
where $k_1$ is the reduced wavenumber and the direction 1 is the direction in which the spectrum is reduced, $\mathrm{Im}$ means the imaginary part, and ${B_2}({k_1}){\cdot}{B_3}^{\ast}({k_1})$ is an element in the energy spectral tensor. The normalized reduced magnetic helicity is defined as $\sigma_{k_1}=k_1{H}_m^r({k_1})/P_{k_1}$, which takes values between $[-1$, $1]$, where $P_{k_1}=|{\bm B}({k_1})|^2$ is the total magnetic energy at wavenumber $k_1$. In practice we first calculate the magnetic helicity as a function of frequency instead of wavenumber, i.e., $\sigma_f={2\mathrm{Im}[B_y(f){\cdot}B_z^*(f)]}/P_f$, where $f$ is the frequency associated with the spacecraft time series of measured magnetic field in the GSE coordinate system. The calculation of $\sigma_f$ is allowed when Taylor frozen-in-flow hypothesis holds, by which the frequency is related to wavenumber \cite{tay38p76,mat82p11,hej11p85}.

A statistical examination is conducted for the nature of the solar wind turbulence. The examination is according to the direction of the long axis of magnetic fluctuations with respect to local ${\bm B}_0$. It is expected that the long axis is quasi-perpendicular to ${\bm B}_0$ for KAW turbulence but to be quasi-parallel for whistler wave turbulence based on the linear kinetic theory \cite{hej12p08}. To determine the long axis of the fluctuations, variance analysis technique is adopted to estimate the long-axis direction, which has maximum variance of fluctuations \cite{son98p85}. A large $\theta_{LB}$, the angle between the long axis and local ${\bm B}_0$, could be expected if it is KAW turbulence.

Figure 1 presents an example to illustrate the parameters used in this letter. Figure 1a plots the magnetic energy spectrum ($P_f$) in the frequency domain (for data on 5 June 2005, 01:33:23$-$01:36:43 UT). The spectrum follows the $f^{-5/3}$ law in the low frequency, and steepens with an index about $-2.8$  when the frequency exceeds $0.4$ Hz. With the frequency approaching 2 Hz, the spectrum flattens again possibly due to the instrument noise and/or aliasing. The two vertical dashed lines in Figure 1a indicate the range of interest in this letter. The left line corresponds to a wavenumber $k\rho_p = 0.1$, where $\rho_p=w_{{\perp}p}/\Omega_{p}$ is the proton thermal gyroradius and $\Omega_{p}$ is the proton cyclotron frequency. The right line is chosen accordingly with $P_f > 10^{-3}$ nT$^2$/Hz that is required for signal level much higher than the instrument noise level \cite{lep95p07}. The wavenumber $k$ represents the reduced wavenumber and is calculated by the Taylor frozen-in-flow hypothesis \cite{tay38p76}, ${2\pi{f}}=kV_p$, with which the frequency domain can be converted to the wavenumber domain. Figure 1a is then replotted as Figure 1b, with $P_k$ instead of $P_f$ for the range bounded by the vertical dashed lines in Figure 1a. In this step an averaging operation is performed to show a smoothed spectrum; the running average window runs over $fe^{-0.5} \leq f \leq fe^{0.5}$. A similar averaging is performed for magnetic helicity spectrum, and the smoothed $\sigma_k$ is shown in Figure 1c. The helicity $\sigma_k$ is obviously non-zero at $k\rho_p \sim 1$ in this example. The drop of $\sigma_k$ at $k\rho_p > 1$ could be attributed to the noise and/or aliasing \cite{kle14p38}. It also could be attributed to the greater balance of KAW turbulence or nonlinear effects at smaller scales that will weaken the helicity signature \cite{hej12p86,mar13p62}.

Figure 1d displays the local spectral index $\alpha_k$, which is determined by the energy spectrum over the same frequency range for the averaging. The index $\alpha_k$ is around $-5/3$ within uncertainty when $k\rho_p < 0.25$; the large uncertainty is inherently due to the short time segment of 200 s that leads to an inefficient coverage for the inertial range. For $k\rho_p > 0.25$, the index $\alpha_k$ rapidly decreases to a more negative value, $\sim -2.8$ at $0.6 < k\rho_p < 1.5$. Note that the more negative $\alpha_k$ occurs with helicity signature enhanced as shown in Figure 1c. The presence of the more negative $\alpha_k$ with the higher helicity allows one to speculate that the steep energy spectrum might be partly attributed to the dissipation of the turbulence with enhanced helicity. An examination on the long-axis direction of magnetic fluctuations, Figure 1e, gives $\theta_{LB} > 80^{\circ}$ for $0.6 < k\rho_p < 1.5$, which is in line with the model of KAW turbulence \cite{hej12p08}.

\begin{figure}
\noindent\includegraphics[width=29pc]{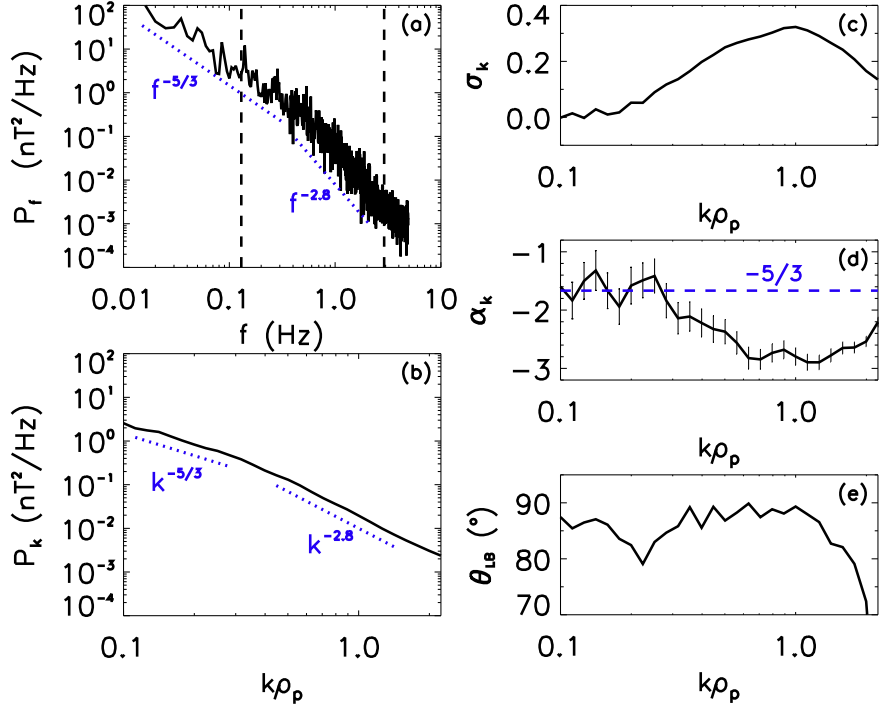}
\caption{An example for a time segment of 200 s on 5 June 2005 to illustrate parameters used in this letter: (a) magnetic energy $P_f$; (b) local average magnetic energy $P_k$; (c) local average magnetic helicity $\sigma_k$; (d) local spectral index of magnetic energy $\alpha_k$; (e) the angle between the long axis of magnetic fluctuations and ${\bm B}_0$, $\theta_{LB}$. Two vertical dashed lines in (a) indicate the range shown in (b)$-$(e) in wavenumber domain.}
\label{fig1}
\end{figure}

\section{Statistical Results}
Statistical investigations are performed to explore the significance of the results in Figure 1. It is found that spectral indices are ordered by the magnetic helicity at proton scales. As an example for $k\rho_p=0.8$, Figure 2 displays, from top to bottom, distribution of $|\sigma_k|$, medians of $\alpha_k$ and $\theta_{LB}$ sorted by $|\sigma_k|$, where $|\sigma_k|$ is the absolute value of $\sigma_k$ that can be either positive or negative even for a specific wave mode, depending on the directions of ${\bm B}_0$ and wave propagation with respect to the Sun. The distribution of $|\sigma_k|$ has a weak peak at $|\sigma_k| \sim 0.22$, and about 61\% of the data have $|\sigma_k| > 0.15$. As $|\sigma_k|$ increases from 0 to 0.4, it is clear that the median of $\alpha_k$ decreases from about $-2.2$ to $-3.2$ (Figure 2b), implying a significant correlation between $\alpha_k$ and $|\sigma_k|$. From Figure 2c, statistical examination concerning the nature of the related turbulence shows that 91\% (63\%) of the data have $\theta_{LB} > 60^{\circ} ~(80^{\circ})$, and there is a tendency of $\theta_{LB}$ increasing with $|\sigma_k|$. If we consider only those data with maximum variances of magnetic fluctuations at least 10 times larger than other variances from orthogonal directions in the examination, the corresponding percentage increases to 98\% (83\%). Such large percentage supports KAW turbulence dominating the nearly collisionless solar wind at proton scales \cite{hej12p08}.

\begin{figure}
\noindent\includegraphics[width=32pc]{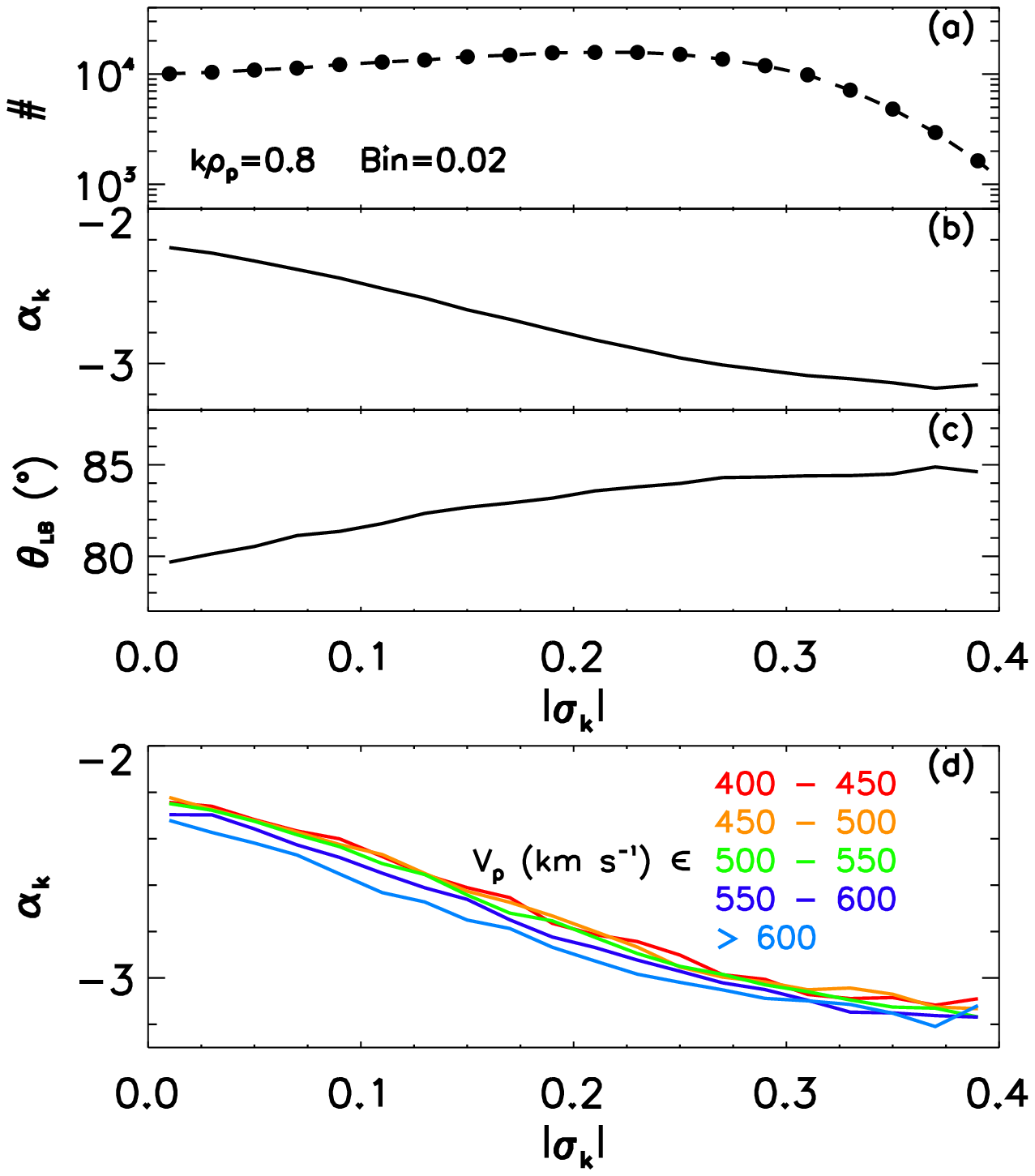}
\caption{Statistics at $k\rho_p=0.8$ for $\sim2.6 \times 10^5$ time segments between 2005 and 2015: (a) distribution of $|\sigma_k|$, (b) median of spectral index $\alpha_k$, (c) median of the angle $\theta_{LB}$, (d) medians of $\alpha_k$, where different color curves are for different ranges of solar wind speed $V_p$.}
\label{fig2}
\end{figure}

\begin{figure}
\noindent\includegraphics[width=29pc]{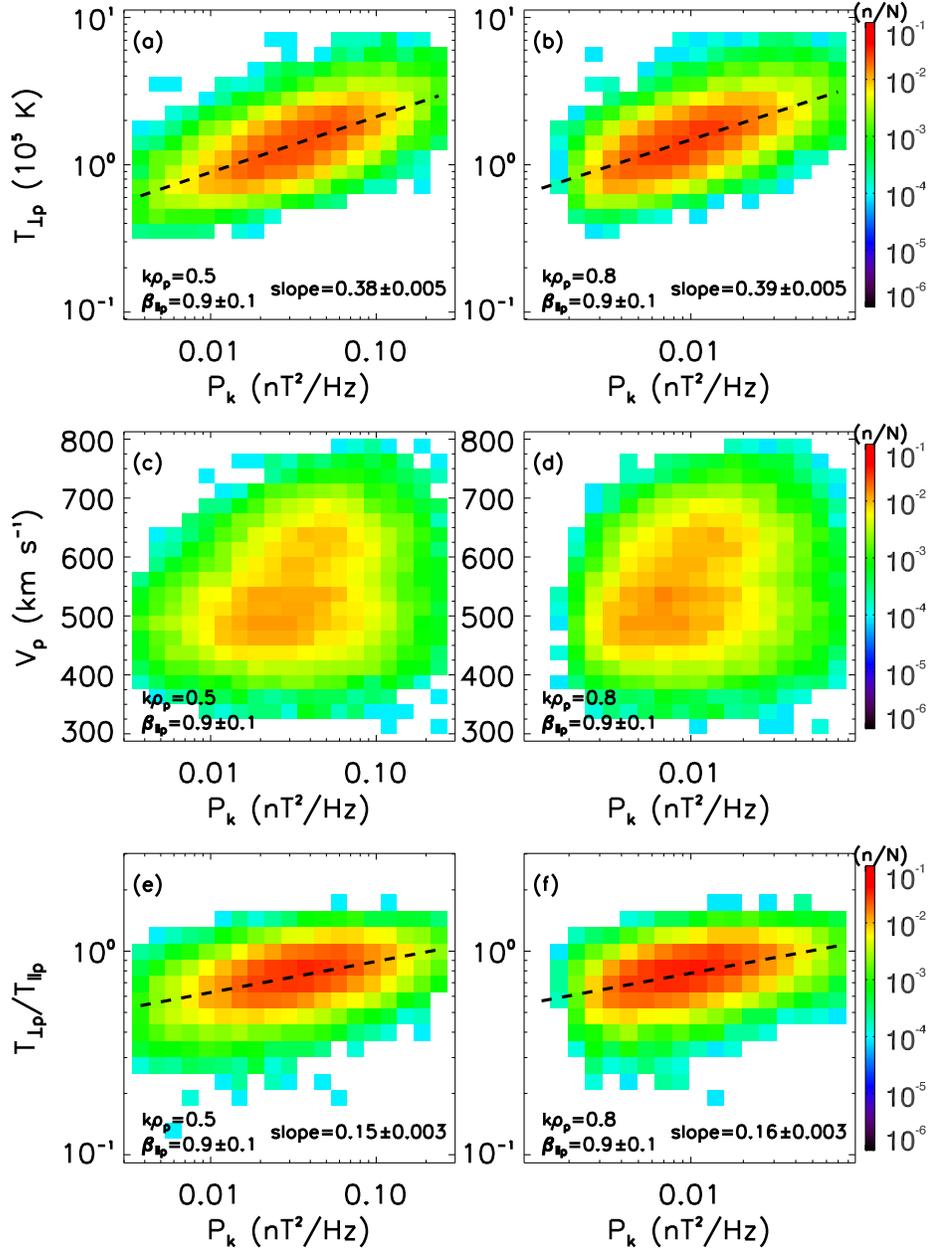}
\caption{Top, middle, and bottom rows are distributions of ($P_k$, $T_{{\perp}p}$), ($P_k$, $V_{p}$), and ($P_k$, $T_{{\perp}p}/T_{{\parallel}p}$), respectively. Left panels are for the wavenumber $k\rho_p=0.5$ while right panels are for $k\rho_p=0.8$. The black dashed lines are the linear fitting of the data in logarithmic space, and $n$ and $N$ represent the data number in each pixel and the total data number in each panel ($\sim 2.3 \times 10^4$), respectively.}
\label{fig3}
\end{figure}

\begin{figure}
\noindent\includegraphics[width=35pc]{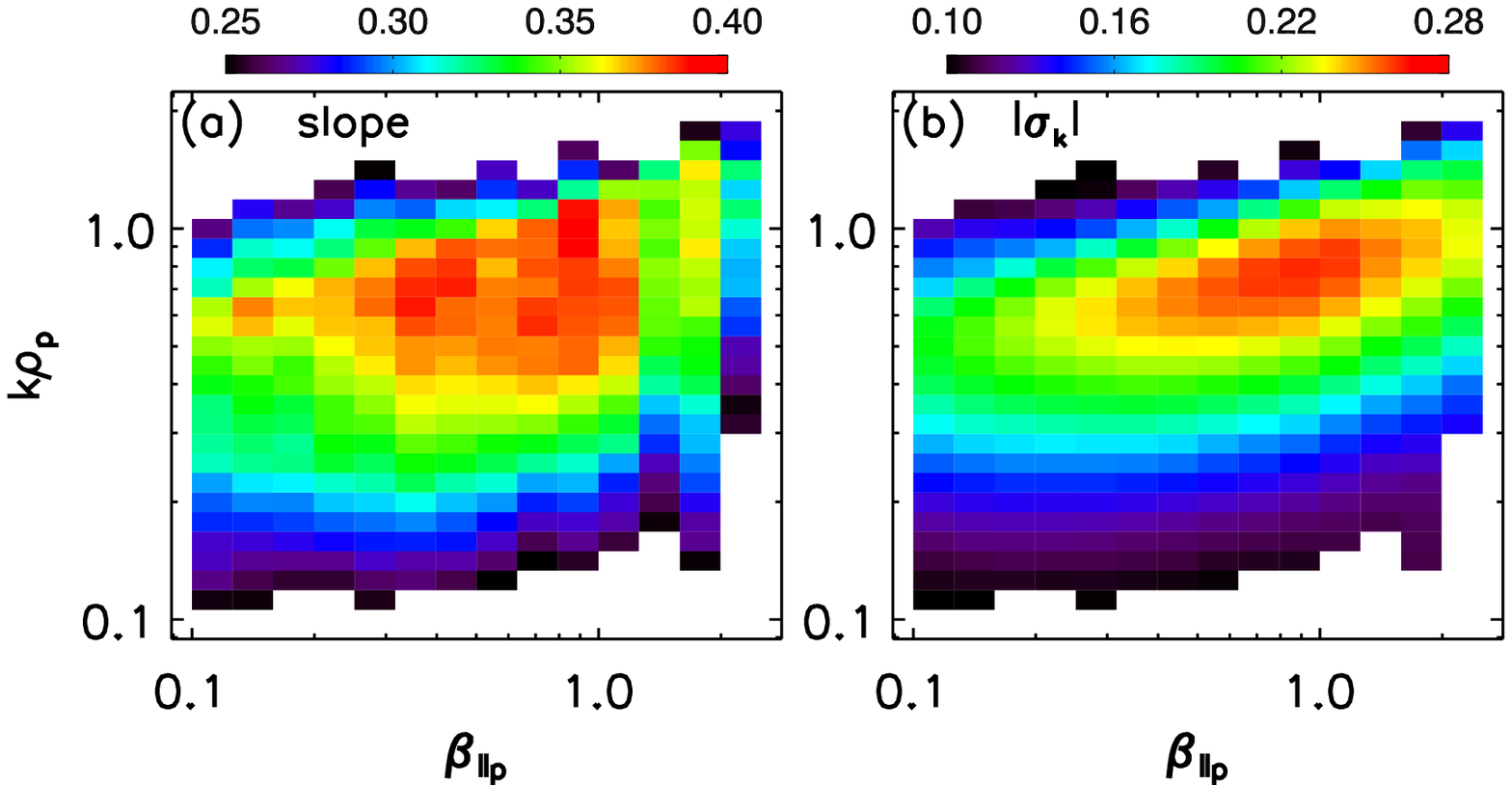}
\caption{Color scale plot for (a) slopes of fitting lines of ($P_k$, $T_{{\perp}p}$) and (b) medians of $|\sigma_k|$ in the ($\beta_{{\parallel}p}$, $k$) space.}
\label{fig4}
\end{figure}

One may speculate that the correlation between $\alpha_k$ and $|\sigma_k|$ is possibly a consequence of the scaling with solar wind speed $V_p$. Figure 2d is presented for this issue. In Figure 2d, different curves are for different sub-datasets bounded by $V_p$, where the red, orange, green, blue, light blue curves correspond to the $V_p$ ranges of 400$-$450, 450$-$500, 500$-$550, 550$-$600, $>$ 600 km s$^{-1}$, respectively. One can see that the correlation exists for each sub-dataset even when the $V_p$ range is narrow (50 km s$^{-1}$). This should imply that the correlation of $\alpha_k$ with $|\sigma_k|$ is common and does not result from the scaling with the solar wind speed.

Another important finding in this letter is that proton temperature $T_p$ is correlated with turbulent energy $P_k$. The correlation is particularly strong for $T_{{\perp}p}$ and depends on $k$, $|\sigma_k|$, and $\beta_{{\parallel}p}$, where $T_p$ and $T_{{\perp}p}$ are the total and perpendicular temperatures of protons, respectively, $\beta_{{\parallel}p}=w_{{\parallel}p}^2/v_A^2$ is the proton parallel beta, $v_A$ is the Alfv\'en speed. Higher $|\sigma_k|$ contributes to a better correlation according to our tests.
Figure 3 plots the distributions of ($P_k$, $T_{{\perp}p}$) at fixed scales of (a) $k\rho_p=0.5$ as well as (b) $k\rho_p=0.8$. All data in the figure are bounded by $|\sigma_k| > 0.15$ and by $0.8 \leq \beta_{{\parallel}p} \leq 1$, producing the data with a size $N \simeq 2.3 \times 10^4$. The black dashed lines present linear fitting of the data in logarithmic space, describing a positive power law for the correlation between $T_{{\perp}p}$ and $P_k$ at the proton scales. The slopes of the fitting lines are considerable, 0.38 at $k\rho_i=0.5$ and 0.39 at $k\rho_i=0.8$. The relative uncertainties are found to be very small, $\sim$ 1\%.

Again, one may keep in mind that the solar wind speed could play a role in the correlation shown in Figure 3. To make this issue clear, Figures 3c and 3d plot the distributions of ($P_k$, $V_{p}$) at fixed scales of $k\rho_p=0.5$ and $k\rho_p=0.8$, respectively. The distributions are shown to be significantly dispersive, although there is a tend of positive correlation between $V_{p}$ and $P_k$. Following the plot of Figures 3a and 3b, five sub-datasets with different $V_{p}$ ranges, as displayed in Figure 2d, are checked. Results show that the correlation between $T_{{\perp}p}$ and $P_k$ remains for each sub-dataset. The slopes of fitting lines decrease relative to those without the limitation of $V_{p}$, but are still considerable. We therefore believe that the correlation between $T_{{\perp}p}$ and $P_k$ is inherent and can not be fully attributed to the scaling with the solar wind speed.

We have also investigated the possible correlation between the temperature ratio $T_{{\perp}p}/T_{{\parallel}p}$ and $P_k$, where $T_{{\parallel}p}$ is the proton parallel temperature. The temperature ratio is a critical parameter in the solar wind data analyses \cite{zha19p75,woo19p53}, as well as in magnetosheath studies \cite{mar18p25}. Figures 3e and 3f present the distributions of ($P_k$, $T_{{\perp}p}/T_{{\parallel}p}$) at scales of (e) $k\rho_p=0.5$ and (f) $k\rho_p=0.8$. One can find a trend that higher $P_k$ corresponds to larger $T_{{\perp}p}/T_{{\parallel}p}$. This implies a correlation between $T_{{\perp}p}/T_{{\parallel}p}$ and $P_k$, whereas this correlation is weaker than that between $T_{{\perp}p}$ and $P_k$. The slopes of fitting lines of ($P_k$, $T_{{\perp}p}/T_{{\parallel}p}$) are around 0.15 in both cases. If we use different datasets by changing the range of $\beta_{{\parallel}p}$ and $V_{p}$, the correlation remains. The slopes of fitting lines may rise when $\beta_{{\parallel}p}$ decreases, but they are usually less than 0.25. A large $T_{{\perp}p}/T_{{\parallel}p}$ may excite proton cyclotron and mirror instabilities, which depends on the proton beta and plays a role in constraining the solar wind plasma \cite{hel06p01}.

It is notable that the correlation between $T_{{\perp}p}$ and $P_k$ with considerable slopes exists in particular at proton scales; at larger scales the correlation weakens greatly. Figure 4a displays the slopes for cases with various $k$ as well as $\beta_{{\parallel}p}$. In each case half of data with higher $|\sigma_k|$ are used to fit the data since higher $|\sigma_k|$ contributes to a better correlation. Any case with a slope $< 0.25$ or relative uncertainty greater than 10\% is marked by the white color. One can see that a highlight region arises at scales $0.3 \lesssim k\rho_p \lesssim 1$ for finite $\beta_{{\parallel}p}$ ($\lesssim 2$), where the slopes are considerably large ($\gtrsim$ 0.35). The slopes depend weakly on $\beta_{{\parallel}p}$ and have a peak value of 0.4 at $\beta_{{\parallel}p} \sim 0.9$. At smaller scales with $k\rho_p > 1$, the result is not clear because at these scales the instrument noise and/or aliasing may be non-negligible \cite{kle14p38}.

As somewhat expected, the large slopes in Figure 4a are accompanied by enhanced magnetic helicity. Using the same data in Figure 4a, Figure 4b plots medians of $|\sigma_k|$ and produces a comparable color region in the ($\beta_{{\parallel}p}$, $k$) space. The comparability between Figures 4a and 4b confirms that higher $|\sigma_k|$ contributes to a better correlation between $T_{{\perp}p}$ and $P_k$. Further investigation on spectral index $\alpha_k$ (not shown) reveals that energy spectra in the highlight regions are steeper and have indices usually lower than $-7/3$, which could be interpreted as hint of the occurrence of turbulent dissipation in the highlight regions. According to Figure 4, on the other hand, there appears an increasing trend of $k\rho_p$ with $\beta_{{\parallel}p}$ for a given slope or $|\sigma_k|$. Such a trend may be related to the increasing trend of $k_b\rho_p$ (dimensionless spectral break position) with $\beta_{{\parallel}p}$, as found by \citeA{dua20p55}.

\section{Summary and Discussion}
This letter performs a statistical research using in-situ measurements of the solar wind lasting for 11 years. For the nearly collisionless solar wind turbulence at proton scales, two new results are obtained as follows. Firstly, the turbulence with a higher magnetic helicity corresponds to a steeper energy spectrum statistically. Secondly, for the turbulence with high helicity ($|\sigma_k| > 0.15$), there exists a positive correlation between the proton perpendicular temperature and the turbulent magnetic energy at scales $0.3 \lesssim k\rho_p \lesssim 1$. The correlation can usually be described by a power law and its slope is considerable ($\gtrsim$ 0.35) for $\beta_{{\parallel}p} \lesssim 2$. In addition, our examination on the nature of the solar wind turbulence supports the model of KAW turbulence at proton scales.

Based on our findings and existing literatures, we propose a scenario for solar wind turbulence and heating as follows. The magnetic fluctuations in the energy-containing range first undergo the Kolmogorov cascade in the inertial range, and predominantly become KAW turbulence at proton scales. The KAW turbulence raises magnetic helicity signature on the one hand, and suffers from dissipation on the other hand. Consequently, part of the magnetic energy goes into protons and a steeper energy spectrum with an index lower than $-7/3$ emerges. During this process, the fluctuations with higher initial energy in the energy-containing range, therefore higher energy at proton scales, will have greater ability to heat protons and result in higher proton temperatures. In other words, the higher temperatures are attributed to higher turbulent energy at proton scales that comes from stronger large-scale magnetic fluctuations by turbulent cascade.

As for the mechanisms of the heating, we propose cyclotron resonance and/or stochastic heating by the proton-scale KAW turbulence. Cyclotron resonance between KAWs and protons is possible, leading to the proton temperature increase in the direction perpendicular to the background magnetic field \cite{lea99p31,gar04p05,hej15p31,ise19p63}. The heating by turbulent KAWs can also be expected when stochastic heating arises due to the presence of large-amplitude electromagnetic fluctuations at the proton gyroscale \cite{joh01p21,cha10p03}. Both mechanisms contribute to the perpendicular heating and are in agreement with our observations. These observations show that the power-law correlation related to the proton perpendicular temperature ($T_{{\perp}p}$) is the best correlation compared with the case for either the parallel temperature ($T_{{\parallel}p}$) or the total temperature ($T_{p}$). The correlation between $T_{p}$ and $P_k$ is secondary with a slope always $< 0.35$, and becomes much weaker between $T_{{\parallel}p}$ and $P_k$. On the other hand, we remind that ion cyclotron waves (ICWs) in the solar wind may also contribute to the proton perpendicular heating. An analysis on the solar wind with time interval of 13.4 hours shows that the distribution of ($\beta_{{\parallel}p}$, $T_{{\perp}p}/T_{{\parallel}p}$) associated with ICWs corresponds to higher $T_{{\perp}p}/T_{{\parallel}p}$ \cite{tel16p79}. Direct measurements of the dissipation rate spectrum around ion scales in magnetosheath turbulence indicate the perpendicular heating by ICWs \cite{hej19p21,hej20p43}. In this letter, we favor the scenario of the heating by KAW turbulence because the KAW turbulence appears to be ubiquitous and has energy much higher than ICWs in the solar wind near 1 AU \cite{hej12p86}. We note that the present scenario is preliminary and more investigations are required in future works.

Finally, we emphasize that the present research is relevant to the frequent observations of high ion perpendicular temperatures in the solar wind and potentially to those in the solar corona \cite{hol02p47,sta19p02}. The data used in this letter are from measurements near 1 AU. Other data, especially produced by the Solar Probe Plus spacecraft in the inner heliosphere \cite{fox16p07}, should be helpful for further studies.

\acknowledgments
The authors acknowledge the SWE team and MFI team on Wind for providing the data, which are publicly available via the Coordinated Data Analysis Web
\url{https://cdaweb.gsfc.nasa.gov/cdaweb/istp_public/}. G.Q.Z. is grateful to the hospitality by Auburn University in USA as a visiting scholar. This research was supported by NSFC under grant Nos. 41874204, 41974197, 41674170, 41531071, 11873018, 41804163, 11903016, and supported partly by scientific projects from Henan Province (19HASTIT020,16B140003).
The authors acknowledge two referees for useful suggestions that improved this paper.


\end{document}